\documentclass[%
reprint,
superscriptaddress,
 amsmath,amssymb,
 aps,
prb,
]{revtex4-2}

\usepackage[dvipsnames]{xcolor}
\usepackage{graphicx}
\usepackage{dcolumn}
\usepackage{bm}
\usepackage{color}
\usepackage{soul} 
\usepackage{changes} 

\begin{document}


\title{Optimal control of molecular spin qudits}

\author{Alberto Castro}
\affiliation{ARAID Foundation, Avenida de Ranillas 1, 50018 Zaragoza, Spain}
\affiliation{Institute for Biocomputation and Physics of Complex Systems (BIFI) of 
the University of Zaragoza, Edificio de Institutos Universitarios de Investigaci{\'{o}}n, Calle Mariano Esquillor, 50018 Zaragoza, Spain}
 \email{acastro@bifi.es}

\author{Adri{\'{a}}n Garc{\'{\i}}a Carrizo}
\affiliation{Institute for Biocomputation and Physics of Complex Systems (BIFI) of 
the University of Zaragoza, Edificio de Institutos Universitarios de Investigaci{\'{o}}n, Calle Mariano Esquillor, 50018 Zaragoza, Spain}

\author{David Zueco}%
\affiliation{Instituto de Nanociencia y Materiales de Arag{\'{o}}n (INMA),
CSIC-Universidad de Zaragoza, Pedro Cerbuna 12, 50009 Zaragoza, Spain}
\affiliation{Departamento de F{\'{\i}}sica de la Materia Condensada, Universidad de Zaragoza,
50009 Zaragoza, Spain}


\author{Fernando Luis}
\affiliation{Instituto de Nanociencia y Materiales de Arag{\'{o}}n (INMA),
CSIC-Universidad de Zaragoza, Pedro Cerbuna 12, 50009 Zaragoza, Spain}
\affiliation{Departamento de F{\'{\i}}sica de la Materia Condensada, Universidad de Zaragoza,
50009 Zaragoza, Spain}

\date{\today}

\begin{abstract}
We demonstrate, numerically, the possibility of manipulating
the spin states of molecular nanomagnets with shaped microwave pulses designed
with quantum optimal control theory techniques. The state-to-state
or full gate transformations can be performed in this way in shorter times than using simple monochromatic resonant
pulses. This enhancement in the operation rates can therefore mitigate the effect of decoherence. The optimization
protocols and their potential for practical implementations are illustrated by simulations performed for a simple
molecular cluster hosting a single Gd$^{3+}$ ion. Its eight accessible levels (corresponding to a total spin
$S=7/2$) allow encoding an $8$-level qudit or a system of three coupled qubits. All necessary gates required for
universal operation can be obtained with optimal pulses using the intrinsic couplings present in this system. The
application of optimal control techniques can facilitate the implementation of quantum technologies based on
molecular spin qudits.  
\end{abstract}

\maketitle


\section{\label{sec:introduction} Introduction}

A crucial challenge for the development of quantum simulation and quantum computation is to scale up computational power while keeping the processor  robust against noise and limiting the complexity of control lines and electronics \cite{Martinis2019,Wu2021,VanDerSypen2021}. A promising strategy is to replace qubits with $d$-dimensional ($d>2$)
quantum systems, or qudits \cite{Gottesman2001,Leuenberger2001,Brennen2005}, as the elementary building blocks of
the quantum architecture. The ability to integrate nontrivial operations in a single physical system helps
simplifying some quantum algorithms \cite{Pirandola2008,Lanyon2009,Kiktenko2015}. In addition, it can also
facilitate their implementation by reducing the number of nonlocal operations, i.e. those connecting different parts
of the circuit.

\begin{figure}
\includegraphics[width=\columnwidth]{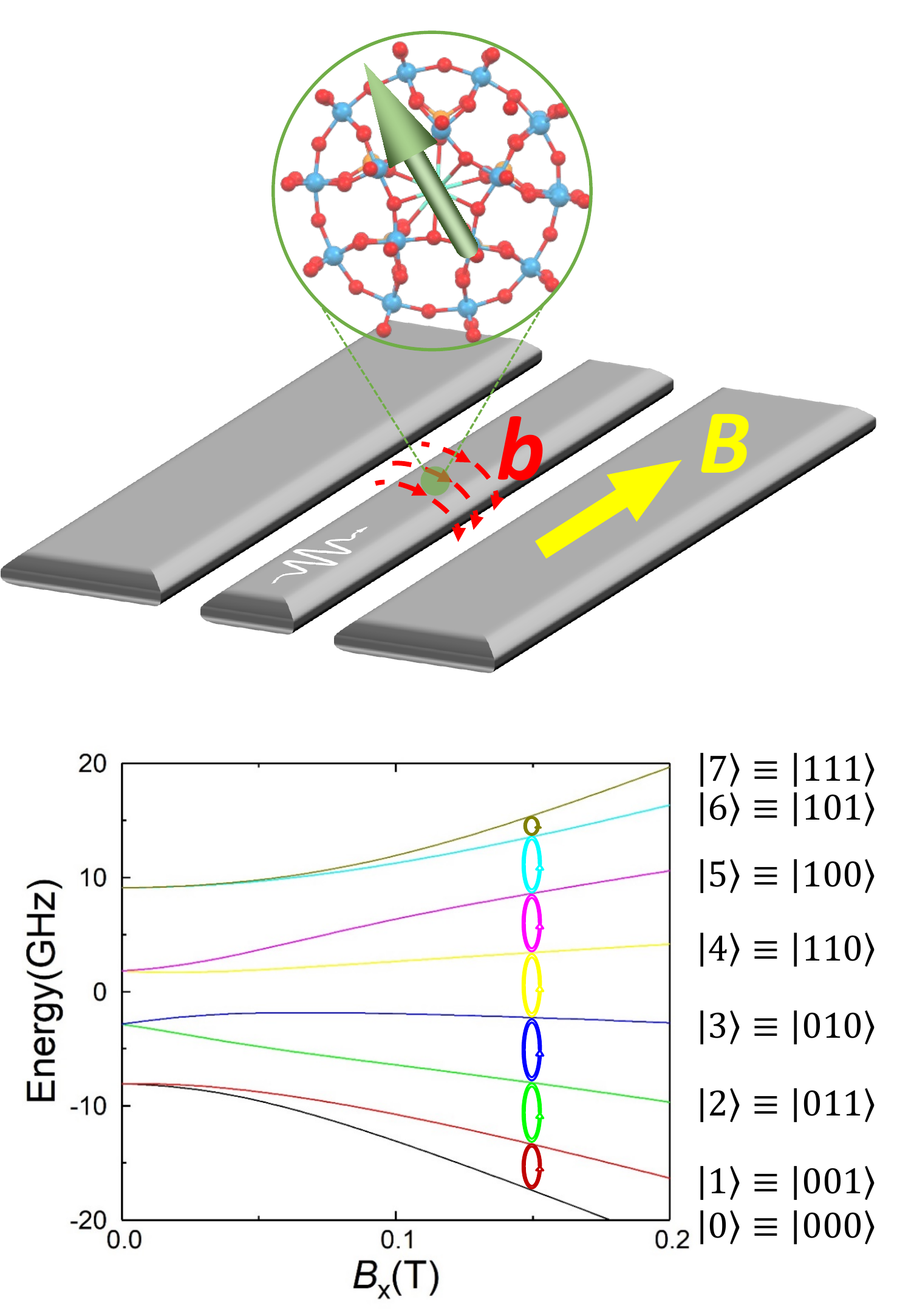}
\caption{\label{fig:scheme}
{\it Top}: Schematic image of a molecular spin subject to a static magnetic field $\vec{B}$ and to an arbitrarily shaped microwave magnetic field $\vec{b}$, here generated by a transmission line. {\it Bottom}: spin energy levels of the GdW$_{30}$ polyoxometalate cluster \cite{martinez-perez2012,Jenkins2017}, whose structure is shown in the top. The $8$ spin states, associated with the $S=7/2$ spin of the central Gd$^{3+}$ ion, enable encoding a $d=8$ qudit or three qubits. The coloured circular arrows mark the seven transitions that can be implemented by means of resonant monochromatic pulses.  
}
\end{figure}

Qudits have been realized with the multiple quantum states of diverse physical systems, including photons \cite{Lapkiewicz2011}, trapped ions \cite{ringbauer2021}, impurity nuclear spins in semiconductors \cite{Asaad2020}
and superconducting circuits \cite{Neeley2009}. Here, we focus on a special class of systems, molecular nanomagnets
\cite{Gatteschi2003,Aromi2012,Atzori2019,Gaita2019,Carretta2021} (see Fig. \ref{fig:scheme} for an illustrative
example). These are coordination or supramolecular complexes that consist of a magnetic core surrounded by a shell
of organic ligand molecules. Chemistry offers nearly unbound possibilities for the design of spin qudits based on
these molecules. The combination of one or several $S > 1/2$ transition metal or lanthanide ions with sufficiently
weak magnetic anisotropy and/or exchange interactions gives rise to a number of low-lying magnetic levels. For
suitably chosen molecular structures, these levels are unequally spaced, making transitions between them addressable
via microwave resonant pulses. And nuclear spin states of the metal ions provide additional resources
\cite{Moreno-Pineda2017,Moreno-Pineda2018,Hussain2018,Gimeno2021,Chicco2021}. 

Examples of molecular based electronic and electronuclear spin qudits, with dimension $d$ ranging from $4$ up to $64$, have been reported recently \cite{Luis2011,Aguila2014,Ferrando2016a,Jenkins2017,Godfrin2017,Moreno-Pineda2017,Moreno-Pineda2018,Hussain2018,Luis2020,Macaluso2020,Gimeno2021,Chicco2021}. In addition, there have been proposals for exploiting their multiple states to specific applications. Relevant examples are the digital quantum simulation of spin-boson models \cite{Tacchino2021}, where the qudit encodes boson states, and the implementation of quantum error correction codes \cite{Chiesa2020,Macaluso2020,Chiesa2021}. The latter is particularly promising, as embedding in each basic unit, in this case a molecule, a suitably designed protection against its specific decoherence sources might represent a huge competitive advantage. Besides, the functionalities need not be defined a priori. When the allowed transitions between different qudit states form a universal set, these systems can be regarded as microscopic size universal processors (or NISQs) \cite{Jenkins2017,Godfrin2017,Godfrin2018,Luis2020,Gimeno2021,Carretta2021}.

However, decoherence remains a serious limitation for fully unleashing the potential of these otherwise very appealing systems. Even though some specifically designed molecular spin qubits show remarkably long coherence times $T_{2}$ \cite{Bader2014,Zadrozny2015a}, decoherence tends to increase for higher spin or higher nuclearity molecules. A sequence of necessarily imperfect gates might become impractically long as compared to $T_{2}$, leading to large error rates. In the paradigmatic example of a qudit-based quantum error protocol, such effects can completely overcome the gain brought about by the code itself \cite{Chiesa2020,Macaluso2020}. Clearly, this calls for more efficient methods to carry out such operations.  

In this work, we consider the possibility of applying quantum optimal control theory (QOCT)
techniques ~\cite{Brif2010} to mitigate some of the limitations associated with the use of 
monochromatic resonant pulses. This theory allows designing more complex pulses in order to 
find the temporal shape of the external perturbation that makes the evolution operator equal
to a predefined gate, as shown for the first time by Palao and Kosloff~\cite{Palao2002}. 
Numerous later calculations employing similar schemes have been 
reported~\cite{Brion2006,Schirmer2009,Reich2012,Hou2014,Chou2015,Arai2015,Dong2016}. We 
illustrate its application to the efficient control of molecular spin qudits by performing 
numerical simulations on a $d=8$ qudit encoded in the $S=7/2$ spin states of a simple 
molecular cluster hosting a Gd$^{3+}$ ion (Fig. \ref{fig:scheme}) \cite{martinez-perez2012}, for which a universal set of operations was realized experimentally \cite{Jenkins2017}.  We have 
parameterized the pulses using a range of different frequencies, and limiting the amplitude 
of each component (since otherwise the optimizations tend to favour high
intensity solutions, that may be experimentally inaccessible, and
moreover, would reduce the coherence times via the excitation
of unwanted levels). The purpose is to assess the expected
gains of using a multi-frequency setup, compared to the usual route to
gate construction through simple monochromatic rotations.

The manuscript is organized as follows. In Section~\ref{sec:methodology} we introduce the
spin Hamiltonian describing the qudit, discuss standard control techniques based on
monochromatic resonant pulses and its limitations in terms of operations speeds and or
fidelities, and describe the optimization methodology used in this work.
Section~\ref{sec:results} shows results obtained by these two methods and discusses their
differences. Finally, Section~\ref{sec:conclusions} summarizes the conclusions and future prospects derived from these results.

\section{\label{sec:methodology} Model and methodology}
\subsection{Spin Hamiltonian: definition of qudit basis states}
We consider a situation like that shown schematically in Fig. \ref{fig:scheme}, where a
molecular spin qudit is tuned by an external magnetic field $\vec{B}$ and controlled by a
time-dependent perturbation able to induce transitions between its different states. The
underlying physics can be described by the following spin Hamiltonian
\begin{equation}
\label{eq:Hamiltonian}
\hat{H} = \hat{H}_{\rm ZF} + \hat{H}_{\rm Zeeman} + \hat{H}_{\rm td}
\end{equation}
\noindent where $\hat{H}_{\rm ZF}$ is the zero-field spin Hamiltonian of the isolated
molecule, $\hat{H}_{\rm Zeeman}$ is the Zeeman interaction with $\vec{B}$ and $\hat{H}_{\rm td}$
is a time-dependent control term. In general, $\hat{H}_{\rm ZF}$ can be quite complex and
include the single ion anisotropy of the magnetic ions forming the molecular core, their
mutual exchange and dipolar interactions as well as the hyperfine couplings to the nuclear
spins. Yet, quite often this Hamiltonian can be simplified. This is the case when dealing
with mononuclear molecules, hosting one metal ion, or when exchange interactions are sufficiently strong to ensure
that only states with the lowest energy total spin value $S$ are significantly populated at the relevant
temperatures. The spin response can then be well 
approximated with the help of a ``giant spin'' approximation~\cite{Gatteschi2003,Wilson2007}. In this work, we have used the following expressions for $\hat{H}_{\rm ZF}$ and $\hat{H}_{\rm Zeeman}$ 
\begin{equation}
\label{eq:zf}
\hat{H}_{\rm ZF} = D\left[ \hat{S}_z - \frac{1}{3}S(S+1)\right]
+ E\left( \hat{S}^2_x - \hat{S}^2_y\right)
\end{equation}
\begin{equation}
\label{eq:Zeeman}
\hat{H}_{\rm Zeeman} = -g\mu_{\rm B}\vec{B}\hat{\vec{S}}
\end{equation}
\noindent where $S$ is the spin quantum number of the molecule, 
$(\hat{S}_x,\hat{S}_y,\hat{S}_z)$ are 
the spin operators, $D$ and $E$ are magnetic anisotropy constants, $g$ is the spin $g$-factor, and $\mu_{\rm B}$ is the Bohr magneton. These expressions describe accurately the GdW$_{30}$ molecular spin qudit,
which we use in section \ref{sec:results} below to illustrate the potential of QOCT techniques. In particular, the uniaxial
magnetic anisotropy $DS_{z}^{2}$ provides the level anharmonicity that is required to properly address individual
transitions between the qudit states. Yet, the methodology is general and could be applied to more complex versions of the
spin Hamiltonian adapted to diverse implementations (e.g. those including weakly coupled electronic spins or a combination
of nuclear and electronic spin states). 

The last term in Eq. (\ref{eq:Hamiltonian}) provides the ability to control the quantum spin states. As with the static terms, for the sake of simplicity we use the following form
\begin{equation}
\label{eq:td}
\hat{H}_{\rm td} = -g\mu_{\rm B}f(t)\vec{b}\hat{\vec{S}}
\end{equation}
\noindent which corresponds to a time-dependent version of the Zeeman interaction term (\ref{eq:Zeeman}). 
This Hamiltonian describes the most common spin control techniques, based on electron
paramagnetic resonance instrumentation, in which the absorption of this signal is the 
crucial  observable. In recent times, it has been shown that spin qubits, including those in molecules, can also
be manipulated by means of electric field pulses \cite{Thiele2014,Liu2019}. In this case, the time-dependent perturbation introduces a
modulation of the crystal field and the magnetic anisotropy terms associated with it. Again, the methods described below are easily adaptable to these situations.

\subsection{Coherent control via monochromatic resonant pulses}
Often, the temporal shape is a simple monochromatic term, e.g.
\begin{equation}
\label{eq:ft}
f(t) = \lambda \cos(\omega t + \phi) \Pi_{t_0}^{t_f}(t)\,,
\end{equation}
where the amplitude is determined by $\lambda$, and $\Pi_{t_0}^{t_f}$ is the rectangular function, that is equal to one
if $t_0 \le t \le t_f$, zero otherwise (in reality, of course, the ramp up and down at $t_0$ and $t_f$ are not abrupt). 
If the frequency is chosen to be close to one of the resonances, and the
amplitude is low enough, the rotating wave approximation (RWA) may be applied, and the effect of these pulses
can be worked out analytically: if $j,k$ are the two levels linked by the resonance (let us assume a perfect resonance, and that
all other frequencies are well separated), the evolution operator $\hat{U}(t)$ is:
\begin{equation}
\hat{U}(t) = \hat{R}_{\vec{n}}^{(jk)}(\theta) \oplus \hat{I}^{(\overline{jk})}\,.
\end{equation}
This expression assumes the interaction representation,
and, in order to simplify the notation, we have set $t_0 = 0$.
The superindex $(jk)$ on the two-dimensional \emph{rotation} operator $\hat{R}_{\vec{n}}^{(jk)}(\theta)$ 
means that it acts on the subspace spanned by the $j,k$ levels, 
whereas the rest of the levels are unaffected ($\hat{I}^{(\overline{jk})}$ is the identity in all but the $j,k$
levels). Within the basis spanned by the two states, $j,k$, and the corresponding Pauli matrices $\sigma_\alpha$ (in that basis),
the rotation operator is given by:
\begin{equation}
\hat{R}_{\vec{n}}(\theta) = \exp\left(-i \frac{\theta}{2} \vec{n}\cdot\hat{\vec{\sigma}} \right)\,.
\end{equation}
$\hat{\vec{\sigma}}$ is the vector of Pauli matrices. The rotation angle $\theta$ 
is $\lambda g \mu_B \vert\mu_{jk}\vert t$,
where $\mu_{jk} = \langle j \vert \vec{b}\hat{\vec{S}}\vert k \rangle$ is the coupling matrix element, and
$\vec{n}$ is the unit vector:
\begin{equation}
\vec{n} = (\cos(\arg \mu_{jk} + \phi), -\sin(\arg\mu_{jk} + \phi), 0)\,.
\end{equation}
The choice of $\phi$ determines the rotation axis: if $\phi = -\arg\mu_{jk}$, we have
a $R_X(\theta)$ rotation; if $\phi = -\arg\mu_{jk} - \pi/2$, we have a $R_Y(\theta)$ rotation. 
We cannot have direct $R_Z(\theta)$ rotations, but they can however be
built by combinations of the former two.
(we recall that here, $X,Y, Z$ do not refer to any spatial direction,
but to the Pauli matrices defined in the basis spanned by the states $j,k$).

Let us consider $R_X(\theta)$ rotations in the following. 
By adjusting the total pulse time $t_f$, one selects the angle $\theta$ and, 
for example, performs a $\pi$-rotation, i.e. if
\begin{equation}
\label{eq:pipulse}
t_f = t_\pi^\lambda = \frac{\pi}{\lambda g\mu_B \vert\mu_{jk}\vert}\,
\end{equation}
the rotation transforms the $j$ state into the $k$ state and viceversa:
\begin{equation}
\hat{R}_X(\pi) = -i \left[\begin{array}{cc} 0 & 1 \\ 1 & 0 \end{array}\right]
\end{equation}
Note the presence of the $-i$ factor; it is an irrelevant global phase factor if we consider the $(j,k)$ subspace
as isolated, but it changes the phase with respect to the rest of the levels.

By concatenation of several of these rotations, one may attempt to construct arbitrary unitaries in
any $2^n$-level system \cite{Jenkins2017,Luis2020,Gimeno2021,Carretta2021}. Some specific quantum gates cannot be constructed in this way, however, due precisely to the presence of the phases mentioned above.
This limitation can be remedied: see~\cite{Kiktenko2015}
for a discussion on this issue, and for an easy solution via the presence
of an extra ancillary level.

The problem that cannot be remedied is the approximate character of all previous expressions, that rely
on the weakness of the perturbation amplitude $\lambda$, thus
avoiding the possibility of arbitrarily speeding up the process by increasing that amplitude.
If we require a minimum fidelity for the $\vert j\rangle \to \vert k\rangle$ transformations,
then a minimum amount of time is necessary.


\subsection{Quantum Optimal Control Theory (QOCT)}

The previous arguments reveal an intrinsic limitation of monochromatic pulses for the
creation of fast quantum gates, which may be further complicated by the
presence of experimental constraints, \emph{i.e.} the inability to 
increase the magnetic field intensities. In order to create
faster gates, one may attempt to use more complex temporal
shapes, \emph{i.e.} combine various frequencies. We therefore wonder
how this possibility may help the control of molecular spin qudits, i.e. whether operation times can be made
substantially shorter than the coherence times. For this purpose, we have applied QOCT. 
We summarize in the following the method and the
basic equations that we have employed.

We consider that the time-dependent pulse-shape function $f(t)$ alluded above
is parameterized with a set of values $u_0, \dots, u_M \equiv u$: i.e. $f = f(u; t)$. 
The evolution of the system is then determined by the Hamiltonian:
\begin{equation}
\hat{H}(u; t) = \hat{H}_0 + f(u; t)\hat{V}\,,
\end{equation}
where $\hat{H}_0 = \hat{H}_{\rm ZF} + \hat{H}_{\rm Zeeman}$, and
$\hat{V} = -g\mu_{\rm B} \hat{\vec{S}} \vec{b}$. 

The evolution operator $\hat{U}(u; t)$ is thus also determined by $u$. In the interaction representation it
evolves according to:
\begin{align}
i \frac{\partial}{\partial t}\hat{U}(u; t) &= f(u, t)\hat{\tilde{V}}(t)U(u; t)\,,
\\
\hat{U}(u; 0) &= \hat{I}\,,
\end{align}
where $\hat{\tilde{V}}(t) = \exp(it\hat{H}_0)\hat{V}\exp(-it\hat{H}_0)$.

The goal is to find a set of parameters $u^{(0)}$ such that the evolution operator 
is equal -- or equivalent -- to a given target gate $\hat{U}_{\rm G}$: $\hat{U}(u^{(0)}, T) = e^{ia}\hat{U}_{\rm G}$
for any irrelevant global phase $a$. In the QOCT framework, this is achieved by finding a set of parameters
that leads to the maximization of a functional of the system evolution; in this case this
can be done by defining this functional as:
\begin{equation}
\label{eq:gatequality}
F(\hat{U}) = \vert \hat{U}\cdot \hat{U}_{\rm target} \vert^2
\end{equation}
where the dot product in the space of linear transformations that we have used is the Fr{\"{o}}benius product:
\begin{equation}
\hat{A}\cdot\hat{B} = \frac{1}{d}{\rm Tr}[\hat{A}^\dagger \hat{B}]\,.
\end{equation}
Here, $d$ is the space dimension. The functional
thus defined acquires its maximum value (one) when $\hat{U}$ is
equal to the target gate, modulo a phase factor. Since, as mentioned above, the evolution is determined by 
the parameters $u$, the problem is reduced to the maximization of the function:
\begin{equation}
G(u) = F\left[\hat{U}(u; t_f) \right]\,.
\end{equation}
Many possible algorithms exist for finding the maxima of multivariate functions such as $G$. Most 
of them require of a means to compute the gradient of the function (in addition to the function itself).
Optimal control theory provides the mathematical tool to derive these gradients
(essentially, Pontryagin's maximum principle~\cite{Pontryagin1962}). 
For the case of our function $G$, the gradient is given by:
\begin{equation}
\label{eq:qoctgradient}
\frac{\partial G}{\partial u_m}(u) = 2{\rm Im}
\int_0^{t_f}\!\!\!{\rm d}t\; \frac{\partial f}{\partial u_m}(u; t)\hat{B}(u; t)\cdot (\hat{\tilde{V}}(t)\hat{U}(u; t))\,,
\end{equation}
where the \emph{costate} $\hat{B}$ is defined by the following equations:
\begin{eqnarray}
i \frac{\partial}{\partial t}\hat{B}(u; t) &= f(u, t)\hat{\tilde{V}}(t)U(u; t)\,,
\\
\hat{B}(u; t_f) &= (\hat{U}_{\rm target}\cdot \hat{U}(t_f) ) \hat{U}_{\rm target}\,.
\end{eqnarray}
Note that it is an equation of motion similar to the one that determines the evolution operator
itself, except the boundary condition is given at the final time of the propagation $t_f$ (it is 
a \emph{final} condition, instead of an initial condition). In consequence, the computation of the
gradient requires two propagations: a forward propagation for $\hat{U}$, and a backward propagation
for $\hat{B}$.

It remains to specify the parameterization of $f$, an important task that actually defines
the set of allowed temporal shapes. This should be done with the experimental capabilities in mind. In 
our case, we have opted for a simple Fourier expansion:
\begin{align}
\nonumber
f(u, t) = \frac{1}{\sqrt{t_f}}u_0 + \sum_{k=1}^{K}
\left[\right.& u_{2k}\frac{2}{\sqrt{t_f}}\cos(\omega_k t) +
\\
\label{eq:fourierexpansion}
&
u_{2k-1}\frac{2}{\sqrt{t_f}}\sin(\omega_k t)\left.\right]\,.
\end{align}
The frequencies $\omega_k = 2\pi k/t_f$, $k=1,\dots, K$ have a maximum \emph{cutoff} value
at $2\pi K/t_f$, that must be chosen big enough to include the relevant natural frequencies of the spin qudit,
but not so large that it cannot be handled experimentally.

Some constraints have to be imposed on the allowed values for the parameters: the pulse amplitude must start
and end at zero: $f(u, 0) = f(u, t_f) = 0$, which translates into $\sum_{k=1}^K u_{2k} = 0$. We
have also imposed a zero value for the average amplitude, $\int_0^{t_f}\!\!{\rm d}t\; f(u, t) = 0$, which
translates into $u_0 = 0$. Finally, the generated magnetic field cannot have arbitrary amplitudes.
Therefore, in the calculations discussed below we have set a maximum value. All these constraints have been added to
the optimization algorithm.

In order to implement these equations, we have used the qutip code as a base~\cite{Johansson2012,Johansson2013}.
We have, however, not employed the QOCT algorithms provided by this platform (at the time of writing of this article),
but used the gradient, computed as in Eq.~(\ref{eq:qoctgradient}), to feed our own QOCT code~\cite{qocttools},
that then utilizes a general purpose function
optimization algorithm: the Sequential Least-Squares Quadratic Programming (SLSQP) algorithm~\cite{Kraft1994}
as implemented in the NLOPT library~\cite{nlopt}.


\section{\label{sec:results} Results}

\subsection{The GdW$_{30}$ molecular spin qudit}

In order to explore the potential of QOCT for the control of spin qudits, we have chosen a system, GdW$_{30}$ \cite{martinez-perez2012,Jenkins2017},
which is both well characterized and relatively simple. The molecular structure of this polyoxometalate cluster is
shown in Fig. \ref{fig:scheme}. It hosts a single Gd$^{3+}$ ion and forms crystals with all molecules oriented in
the same manner. In addition, magnetically diluted crystals can be grown by simply replacing Gd$^{3+}$ with
Y$^{3+}$, which is chemically equivalent but diamagnetic. This allows enhancing spin coherence times up to $2-3$
$\mu$s \cite{martinez-perez2012} while keeping the possibility of orienting the magnetic fields $\vec{B}$ and $\vec{b}$ along specific
molecular axes. This molecule shows a hard magnetic
axis along the main molecular axis $z$. Its static spin Hamiltonian can be well described by Eqs.
(\ref{eq:Hamiltonian}), (\ref{eq:zf}) and (\ref{eq:Zeeman}) with $S = 7/2$, $g=2,$ $D=1281$ MHz and $E=294$ MHz.
The overall splitting of the $d=8$ multiplet is smaller than $1$ K, or $20.8$ GHz, thus ensuring that adjacent
level splittings lie within the reach of conventional EPR as well as of other microwave technologies. In all
calculations discussed below, we have chosen the static magnetic field $\vec{B}$ to point in the $x$ direction
(medium axis) and set $B=0.15$ T. Under such conditions, the eigenstates of $\hat{H}_{\rm ZF} + \hat{H}_{\rm Zeeman}$ become close approximations to pure spin projections along $\vec{B}$. The time-dependent magnetic field $\vec{b}$ is
perpendicular, and points in the $y$ direction (easy axis), thus inducing transitions between adjacent energy
levels. 

\subsection{Transition implementation via monochromatic pulses}
\begin{figure}
\centerline{\includegraphics[width = 0.95\columnwidth]{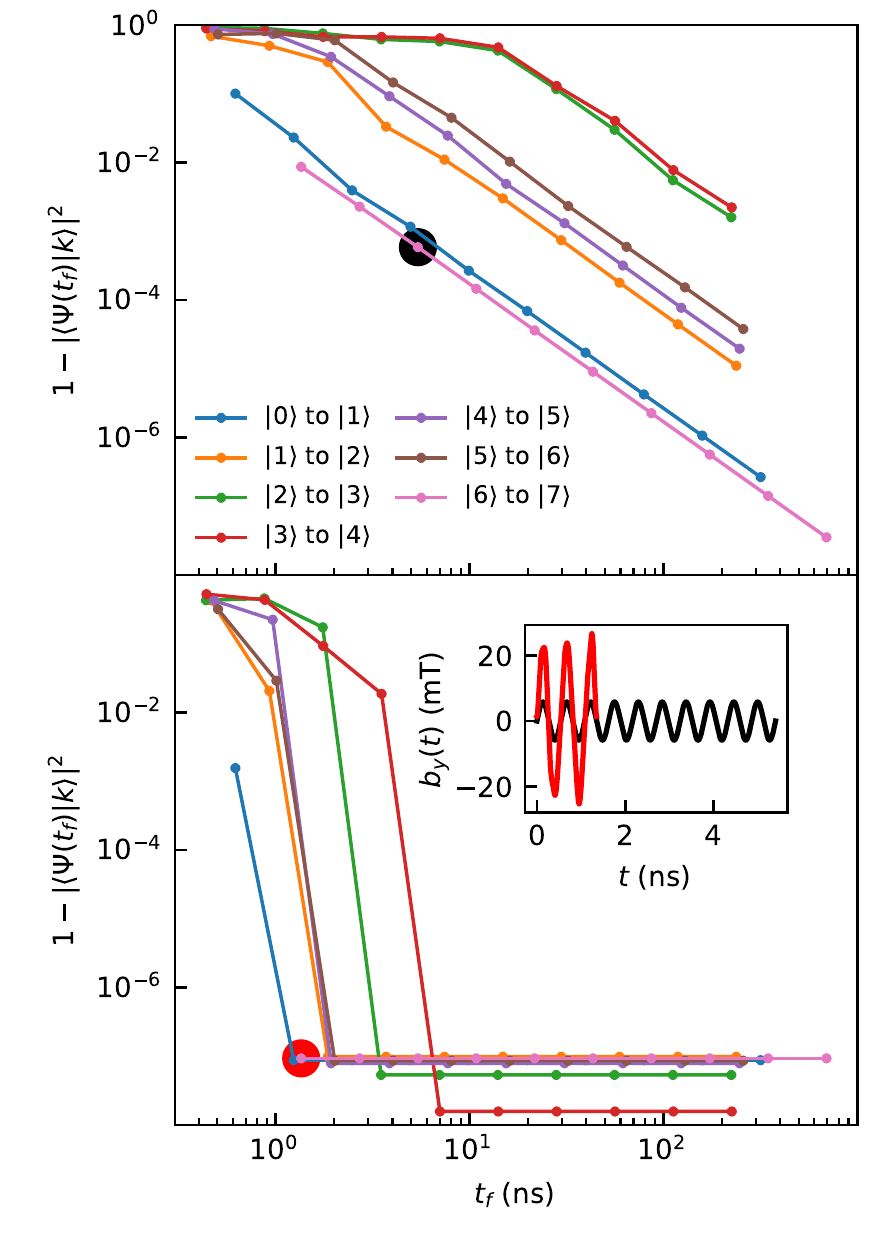}}
\caption{\label{fig:fdt}
{\it Top:} Infidelities of the seven main transitions in the GdW$_{30}$ molecule, as function
of the $\pi$-pulse time. {\it Bottom:}
Infidelities of the seven main transitions in the GdW$_{30}$ molecule, as function
of the total pulse time, for pulses obtained with optimal control. {\it Inset:}
Time-dependent shape of the pulses used to generate two of those $\vert 6\rangle$ to $\vert 7\rangle$
transitions: one $\pi$-pulse (black), and one optimized pulse (red), corresponding
to the thick black and red dots, respectively, on the pink lines.
}
\end{figure}

In this section, we discuss the manipulation of the GdW$_{30}$ spin using monochromatic pulses resonant with the set of allowed transitions
mentioned above and shown in Fig.~\ref{fig:scheme}. As an illustration,
we consider the application of $\pi$ pulses linking every two of these states. Fig.~\ref{fig:fdt} (top)
displays the transformation infidelities (i.e. $1 - \vert\langle\psi(t_\pi^\lambda)\vert k\rangle\vert^2$) as 
a function of the time $t_\pi^\lambda$ allocated to complete the operation, for the seven $\vert j \rangle \to \vert k = j+1\rangle$ transitions.  One can see that the error in the outcome state is reduced as $t_\pi^\lambda \to \infty$. 
In fact, from the logarithmic plot one may infer a quadratic behaviour:
\begin{equation}
1 - \vert\langle\psi(t_\pi^\lambda)\vert k\rangle\vert^2 
= \mathcal{O}((1/t_\pi^\lambda)^2)\,.
\end{equation}
It becomes clear that, in order to ensure a given fidelity, a minimum time (or,
equivalently, a maximum amplitude) is required. 


\subsection{Optimization of state-to-state transitions}

In this and the following section, we apply the QOCT methods described in section \ref{sec:methodology} to quantum operations on GdW$_{30}$ having different targets. First, we optimize resonant transitions and sequences of these. Then, we tackle the optimization of quantum gates. The goal here is 
to see how OCT permits to increase the fidelities shown in Fig.~\ref{fig:fdt} (top), by allowing
for the presence in the pulse of other frequencies, besides the resonant one. We have performed
QOCT calculations~\footnote{
Note that the OCT for the simpler problem of state population -- in contrast to the the harder
problem of the creation of a given evolution operator or gate -- requires equations
that are slightly different to the ones described in Section~\ref{sec:methodology}.} 
considering, for each transition, the same total propagation time used to create Fig.~\ref{fig:fdt} (top).
Each of these propagation times $t_{\pi}^{\lambda}$ corresponds to a $\pi$-pulse amplitude $\lambda$ [Eq.~(\ref{eq:pipulse})], 
that we have used now
to set a bound for the Fourier expansion coefficients
in the OCT maximization: the temporal shape of the microwave field
is given by Eq.~(\ref{eq:fourierexpansion}), where $\vert \frac{2 u_j}{\sqrt{t_{\pi}^{\lambda}}}\vert \le \lambda$. Having in mind the level splitting present in GdW$_{30}$ and typical experimental capabilities, we have set the frequency cutoff at 8 GHz. 
The resulting (in)-fidelities are displayed in Fig.~\ref{fig:fdt} (bottom). The
shaped pulses permit to decrease those infidelities with respect
to the simple $\pi-$pulse values, down to negligible values 
for all but the shortest total transition times
(we have set a 10$^{-7}$ threshold to stop the search algorithm, and hence the flat
curves for the longer times). In Fig.~\ref{fig:fdt} (inset), we also compare the shape of control pulses corresponding to a resonant transition (monochromatic $\pi$-pulse) and to the optimized one from states $\vert 6\rangle$ to $\vert 7\rangle$. These examples correspond to duration values marked by (respectively) black (top) and red dots (bottom). It is clear that the optimized pulse achieves a much better fidelity in a much shorter time.


\begin{figure}
\centerline{\includegraphics{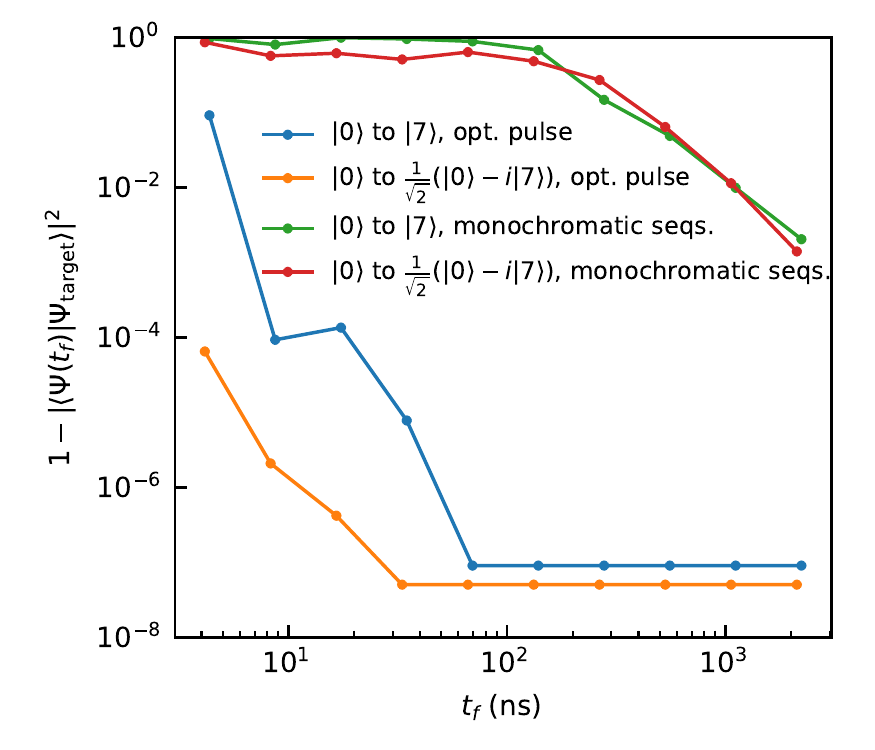}}
\caption{\label{fig:0to7}
Infidelities for the $\vert0\rangle\to\vert7\rangle$ and
$\vert0\rangle\to \frac{1}{\sqrt{2}}(\vert 0\rangle -i \vert7\rangle)$ transitions, as a function
of operation time, for the pulses obtained as a sequence of monochromatic resonant pulses ($\pi$ or $\pi/2$), and
for the optimal pulses obtained with QOCT.}
\end{figure}

Let us consider now the more general case of optimizing transitions between states, say $| n \rangle $ and $|m \rangle$,  that are not directly coupled by the external field, i.e.  $\langle n | S_y | m \rangle = 0$.   %
A possible solution is to concatenate a series of $\pi$-pulses, 
between intermediate states. In fact, this is a criterion for universality:
if any two states can be connected through others, the qudit can perform any unitary and, in this sense, can be 
regarded as a universal quantum processor. However, the time needed is proportional to the number of pulses, which in
practice is a  limitation. Thus, in this case, QOCT offers a clear advantage by replacing a single shaped pulse to
achieve, and accelerate, a process that would otherwise require a sequence of $7$ monochromatic pulses.  

In Fig.~\ref{fig:0to7} we have tested this by comparing the infidelity in the transition $|0\rangle \to |7\rangle$,
using a sequence of $\pi$-pulses between adjacent levels, $|k \rangle \to |k + 1 \rangle$, and using QOCT. 
The improvement is quite significant. In particular,
we have run over a range of amplitudes $\lambda$, that determine the $\pi$-pulse
length for each transition, $t_\pi^{\lambda}(k\to k+1)$. The full 
$\vert 0\rangle\to \vert 7\rangle$ process then requires 
$t_f = \sum_{k=0}^{6}t_\pi^{\lambda}(k\to k+1)$. The plot displays the fidelity
achieved with these pulse sequences. Then, for each of those times, we have
performed QOCT calculations, once again setting a bound for the amplitudes of the individual
Fourier terms equal to $\lambda$. The plot shows how, even at
very short operation times, the fidelities achieved by the optimized pulses
are almost equal to one. In terms of time scales, a $99$ \% fidelity can be achieved in less than $10$ ns, thus
much shorter than $T_{2} \simeq 2$ $\mu$s, while reaching the same result with a sequence of monochromatic pulses
would take more than $1$ $\mu$s.

In Fig~\ref{fig:0to7} we also show results of a similar calculation, but using
the $\vert0\rangle\to \frac{1}{\sqrt{2}}(\vert 0\rangle -i \vert7\rangle)$ state
as target, a superposition state that can be reached with a $\pi/2$-pulse corresponding
to the $\vert 0\rangle \to \vert 1\rangle$ transition, followed by the same previous sequence
of $\pi$-pulses that raises the state through the next adjacent levels. The results
are qualitatively similar to the ones obtained for the full $\vert 0\rangle$ to $\vert 7\rangle$ transition, thus showing that the speed
enhancement achieved by the application of QOCT is not restricted to any
particular class of transitions. This allows targeting the optimization of complex gates, which is discussed next. 


\begin{figure}
\includegraphics[width=\columnwidth]{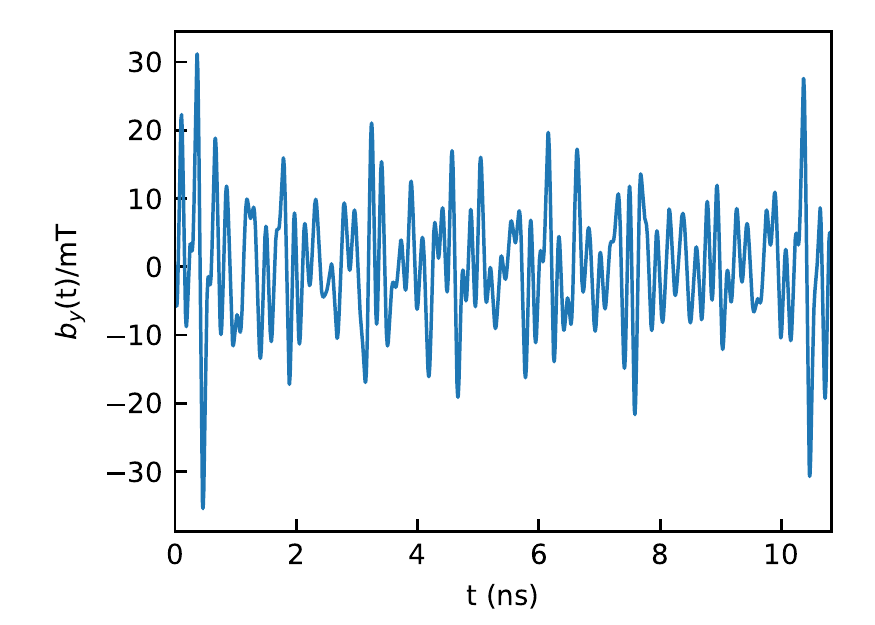}
\includegraphics[width=\columnwidth]{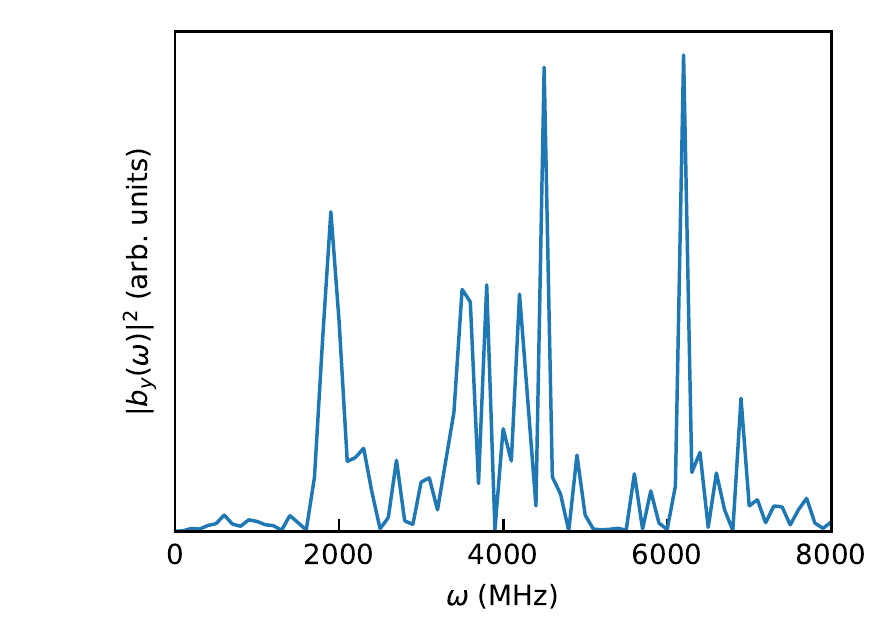}
\caption{\label{fig:ftw2}
Optimal pulse in the time domain (top), and its power
spectrum in the frequency domain (bottom), of the optimal
pulse obtained for the $\theta = \pi/4$ Deutsch gate.
}
\end{figure}


\subsection{Quantum gate optimization}

Finally, we proceed to our true objective: the search for non-trivial pulse shapes that
realize quantum gates, with high fidelities, in short times. As target
gates, we have chosen the family of Deutsch gates~\cite{Deutsch1989,XiaoFeng2018}:
\begin{equation}
D(\theta) = \left[\begin{array}{ccc}
I_6 & 0 & 0
\\
0 & i\cos(\theta) & \sin(\theta)
\\
0 & \sin(\theta) & i\cos(\theta)
\end{array}\right]\,,
\end{equation}
where $I_6$ is the $6 \times 6$ identity matrix. Note that this family includes
the Toffoli gate, for $\theta = \pi/2$. The reason for focusing on
this set of gates is that it is universal: any circuit
can be constructed by combination of these components.

The total propagation time $t_{\rm f}$ has been set to $20$ times the
maximum natural period of the field-free Hamiltonian, i.e. the period
corresponding to the smallest transition frequency. For the choice of 
static magnetic field used here ($\vec{B} = 0.15 T \vec{e}_x$),
$t_{\rm f} \sim 10$ ns.

As discussed above, the optimizations are performed
constraining the allowed parameter set, such that each sinusoidal (or cosinusoidal)
term in the expansion (\ref{eq:fourierexpansion}) has a maximum amplitude, i.e.
\begin{equation}
\vert\frac{2u_k}{\sqrt{t_f}}\vert b \le b_{\rm max}\,.
\end{equation}

In the calculations shown here, we have set $b_{\rm max}$ to a relatively high value of $20$ mT in order to make the operation, and therefore also the computational, times manageably short. The influence of $b_{\rm max}$ on $t_{\rm f}$ is discussed below. The optimization algorithm is an iterative process that we stop when
the quality of the gate, measured as $F(\hat{U}(u; t_f)) = \vert \hat{U}(u, t_f)\cdot \hat{D}(\theta)\vert^2$,
reaches a certain threshold, that for these calculations we have set to $0.99$.

Figure~\ref{fig:ftw2} displays the results obtained for $\theta = \pi/4$ as an example (the results
obtained for other angles are qualitatively similar).
The top panel shows $f(u^{(0)}; t)$ in real time, whereas
the bottom panels displays its power spectrum. Both plots demonstrate
the complexity of the pulses, that do not have dominant frequencies.

\begin{figure}
\includegraphics[width=\columnwidth]{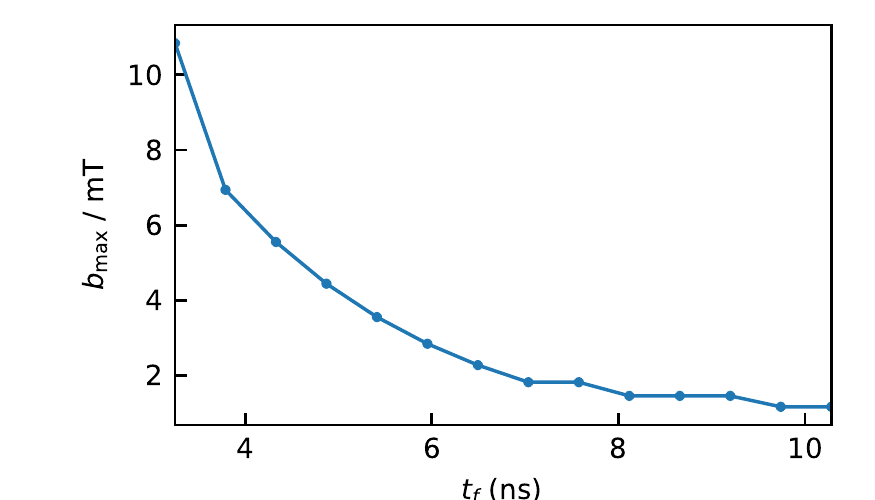}
\caption{\label{fig:hmax-T}
Minimal bound on the amplitudes ($y$-axis) that allows 
to obtain a $\theta = \pi/2$ Deutsch gate for a given total operation time ($x$-axis).
}
\end{figure}

In the previous example, we have set a total operation time $t_{\rm f}$ and
an amplitude bound $b_{\rm max}$. Those magnitudes are of course related:
if, for a given $t_f$, we set a too low amplitude bound, the quality of the gate (as measured 
by Eq.~(\ref{eq:gatequality}) will also be too low. In fact, if we fix a threshold for acceptable
gate quality (say, $0.99$), for each given propagation time there will be a minimum
amplitude bound necessary for the QOCT algorithm to return a successful pulse.
Even with optimal pulse shapes, we need a minimum of field amplitudes in order
to get a high quality gate. We have therefore studied this issue, computing the minimum
amplitude bound that can be used to constrain the QOCT calculation in order to
get a given gate, as a function of operation time. The results are shown in Fig.~\ref{fig:hmax-T}.

This type of plot helps to ascertain whether or not the gate operations are experimentally feasible,
as in practice there is a hardware bound on the field amplitudes that can be used. Given
this limit, one may learn from the plot what operation times are feasible, even with shaped
optimal pulses. Obviously, the lower the available amplitudes, the longer the operations times must be.

\section{\label{sec:conclusions} Conclusions}

The previous results show that the use of complex pulses, engineered by optimal control techniques, provides a method for improving the speed of operations performed on spin qudits. The advantages are already noticeable for
the realization of elementary transitions, when the application of monochromatic pulses is limited by the need of
keeping the excitation amplitudes sufficiently low. Yet, they become even more important when dealing with more
complex operations. Then, QOCT allows reaching any given fidelity of the outcome wave function with a single
control pulse replacing the often long sequence of resonant pulses required by standard techniques. This
possibility is especially relevant for algorithms that involve transitions between relatively disconnected states.
For a necessarily limited coherence spin time, this difference can represent a big gain in the performance of such protocols.

The molecular qudit design and the way information is encoded on its spin
states can be adapted to suit best the requirements of specific
quantum protocols \cite{Carretta2021,Petiziol2021}, thus offering a vast
choice of possible molecular platforms and implementations. Optimal control
techniques described in this work are flexible enough to be made compatible with almost any of them. Although we have here considered a
specific molecule for illustrative purposes, QOCT can deal with any spin model and with diverse control
interactions (e.g. magnetic or electric field pulses). Therefore, it can be adapted to boost the implementation of
diverse algorithms. We feel that it will be of special relevance to quantum error correction, because reaching a
fidelity improvement with such protocols critically depends on the ratio between the implementation time and
$T_{2}$. With an additional computational cost, one can even consider optimizing the control pulses to
best compensate for the actual sources of decoherence in each molecule, mainly dephasing by nuclear spins located
in the ligand shell surrounding the magnetic core.

The use of more sophisticated control pulses represents a
challenge to experimental implementations. Most commercial
EPR systems work with relatively
narrow excitation bands and a reduced choice of pulse shapes.
In recent years such systems have been complemented with
waveguide generators able to arbitrarily design the
excitation pulses \cite{Tseitlin2011,Spindler2016,Prisner2019}. Still, these set-ups are limited to
frequencies lying sufficiently close to the cavity resonance
frequency. In order to expand the frequency window, one can
resort to on-chip circuits, with the excitation being driven
by an open transmission line. This scheme, illustrated by Fig. \ref{fig:scheme}, has been applied to
investigate coherent control of NV centers in diamond \cite{DeLange2010} and
to perform broadband spectroscopy of GdW$_{30}$ and other molecular spin
qudits \cite{Jenkins2017,Gimeno2021}. It can also be
used to read out the outcome, either by looking at the frequency
dependent absorption, in a projective measurement, or by
coupling it to a superconducting resonator that can perform nondemolition, dispersive measurements of the qudit
states \cite{gomezleon2021}. Some experimental systems combining
superconducting resonators and broadband control lines have been recently reported \cite{Keyser2020}. Another promising
implementation is based on the combination of single-molecule electronics with
gates or coils able to locally generate arbitrarily shaped electric or
magnetic microwave pulses. Experiments performed on molecules trapped between
point contacts or between a metal substrate and a STM tip have provided the
first measurements of spin coherence in individual molecules \cite{Zhang2021}
and achieved the realization of Grover's search algorithm using three nuclear
spin states in a Tb-based molecule \cite{Godfrin2017}. 

In summary, the application of quantum optimal control theory to operate the states of molecular spin qudits
offers remarkable prospects to improve their performance, compensating for their not too long coherence times.
Equipped with these techniques, many more molecular systems and applications can become feasible, thus
contributing to an alternative and promising path towards large scale quantum computation and simulation. 

\begin{acknowledgments}
We acknowledge the financial support from grants FIS2017-82426-P, PGC2018-094792-B-I00, RTI2018-096075-B-C21 and PCI2018-093116 funded by MCIN/AEI/ 10.13039/501100011033 and “ERDF A way of making Europe”, 
grant PID2020-115221GB-C41/AEI/10.13039/501100011033, the European Union’s
Horizon 2020 research and innovation programme (QUANTERA project SUMO, FET-OPEN grant 862893 FATMOLS), the Gobierno
de Arag\'on grant E09-17R-Q-MAD and the CSIC Quantum Technology Platform PT-001.
\end{acknowledgments}




\bibliography{GdW30gates}

\end{document}